\documentclass[aps,pre,groupedaddress,twocolumn]{revtex4}
\pdfoutput=1

\usepackage{hyperref}
\usepackage{subfigure}
\usepackage{color}
\usepackage{graphicx}

\begin{document}
\expandafter\ifx\csname urlprefix\endcsname\relax\def\urlprefix{URL }\fi

\DeclareGraphicsExtensions{.pdf, .jpg}

\title{\large
   {\it N}-slit  interference:  fractals  in  near-field
   region, \\ bohmian  trajectories.  }

\large
\author{Valeriy I. Sbitnev}
\email{valery.sbitnev@gmail.com}
\affiliation{B.P.Konstantinov St.-Petersburg Nuclear Physics Institute, Russ. Ac. Sci.,
     Gatchina, Leningrad district, 188350, Russia.}


\date{\today}

\begin{abstract}
Scattering cold particles on an $N$-slit grating is shown to reproduce an interference pattern, that manifests itself in the near-field region as the fractal Talbot carpet. In the far-field region the pattern is transformed to an ordinary diffraction, where principal beams are partitioned from each other by ($N-2$) weak ones. A probability density plot of the wave function, to be represented by a gaussian wavepacket, is calculated both in the near-field region and in the far-field one. Bohmian (geodesic) trajectories, to be calculated by a guidance equation, are superimposed on the probability density plot well enough. It means, that a particle, moving from a source to a detector, passes across the grating along a single bohmian trajectory through-passing one and only one slit.\\


\end{abstract}

\maketitle

\large

\section{\label{sec:level1}Introduction.}

 Wave interference is a most impressive phenomenon be it induced by waves on water, acoustic waves, or electromagnetic ones -  radio-waves, light waves,  $\gamma$-radiation. Quantum-mechanical experiments  dealing with interference phenomena~\cite{Cronin:2007} display interesting
 interference phenomena. They involve self-affine fractal quantum evolution of the probability densities and quantum revivals~\cite{BerryKlein:1996, Berry:1996, Berry:2001, AmanatidisEtAl:2003, Sanz:2005}.
 There is a special interest to the phenomena from the side of quantum computation and communication~\cite{ClauserDowling:1996, Steane:1998, OlmschenkEtAl^2009}.

\begin{figure}[htb!]
  \centering
  \begin{picture}(100,170)(120,150)
      \includegraphics[scale=0.55]{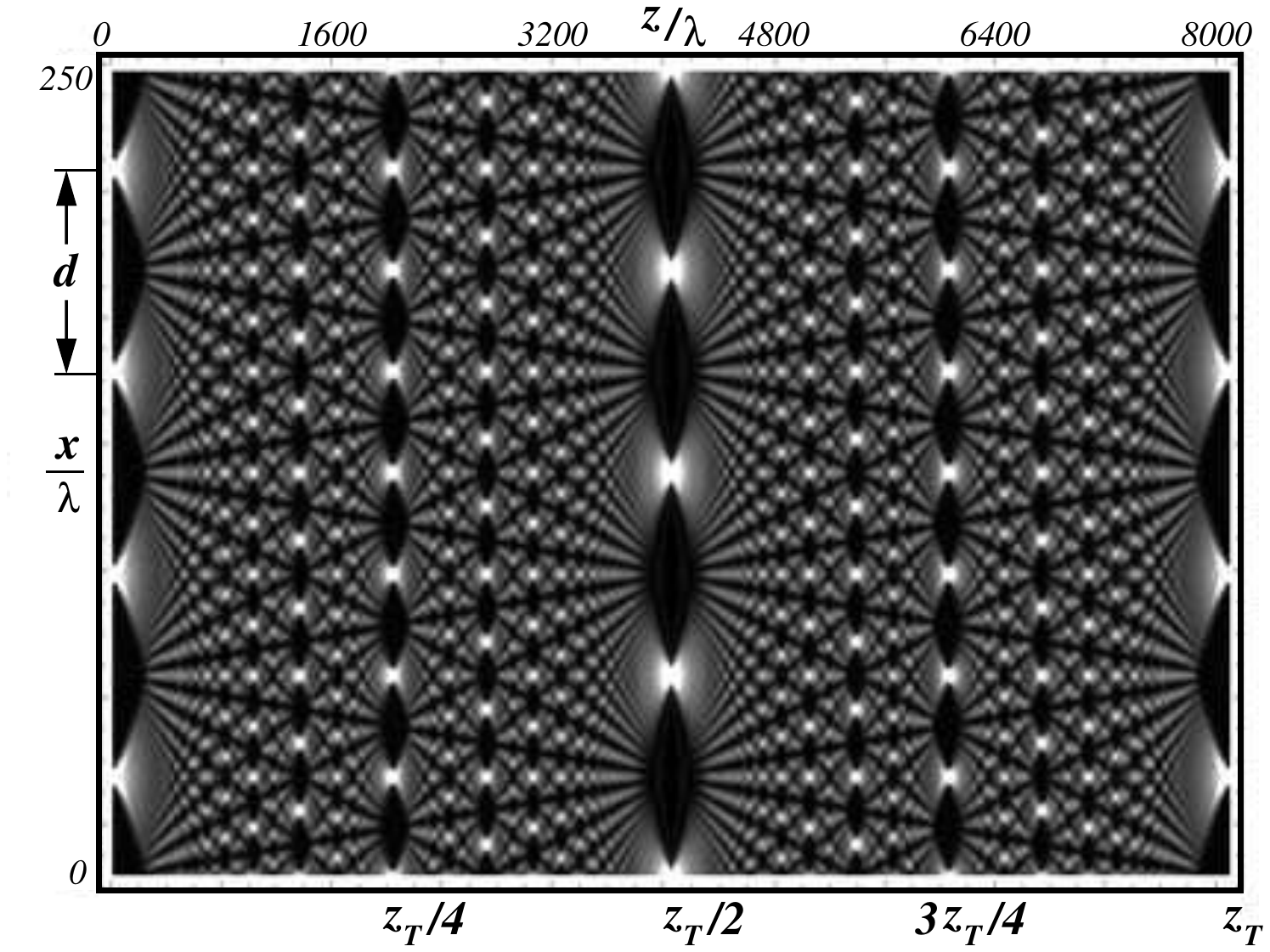}
  \end{picture}
  \caption{
 The optical Talbot effect for monochromatic light, shown as a "Talbot Carpet". The figure has been captured from
 ~{http://en.wikipedia.org/wiki/Talbot{\_}effect}.
 }
  \label{fig=1}
\end{figure}

  Most fantastic interference effects are disclosed in near-field region, i.e., in the vicinity of an interference grating. In this region complex wave interference shows very exotic patterns named in literature as the Talbot carpets~\cite{BerryKlein:1996, Berry:1996, Berry:2001}. They disclose fractal-like self-similar structures. Henry Fox Talbot was first who observed in 1836 such an effect in a near-field region\footnote[1]{http://en.wikipedia.org/wiki/Talbot{\_}effect}.
  Fig.~\ref{fig=1}, for example, demonstrates the optical Talbot carpet for monochromatic light to be scattered on 4-slit grating.
 A significant parameter in this optical pattern is the Talbot Length
 \begin{equation}\label{eq=1}
    z_{\;\rm T} = {{2d^{\,2}}\over{\lambda}},
 \end{equation}
 where $d$ is the period of the diffraction grating and $\lambda$ is the wavelength of the light incident on the grating.
 We will deal with this parameter often enough at representations of interference patterns in near-field regions.
 Instead of monochromatic light scattering, however, we will simulate here particles (cold neutrons) scattering on nanoscale gratings
 (the period $d$ is multiple of the particle wavelength).

\begin{figure}[htb!]
  \centering
  \begin{picture}(100,100)(170,280)
      \includegraphics[scale=0.75]{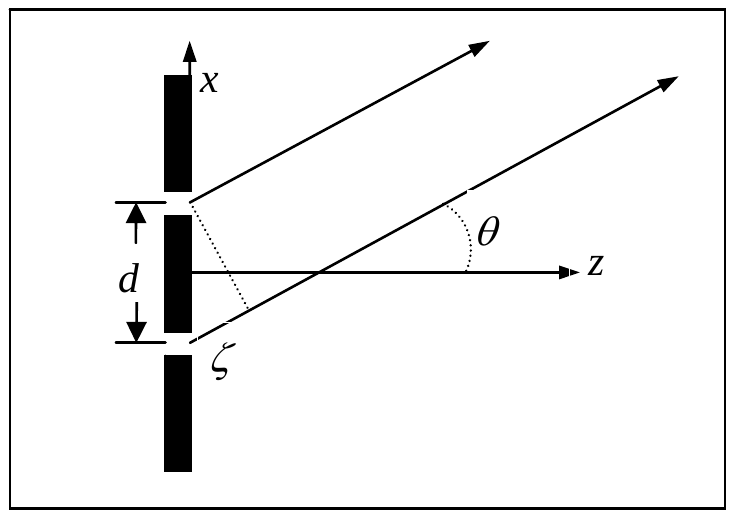}
  \end{picture}
  \caption{
 Diffraction from two slits in far-field region: $d$ is distance between the slits, $\theta$ is  deflection angle with respect to axis $z$,
 and $\zeta$ is phase shift of wave rays emitted from the slits.
 }
  \label{fig=2}
\end{figure}

 On the other hand,
 as a distance from the slit grating to a detector increases the exotic near-field interference transforms to a diffraction pattern observed in the far-field region. Beautiful magnificence of the fractal structures disappears. Instead of it divergent rays from the slit source come into being.
 Intensity of these rays is described by a well-known formula~\cite{Sbitnev:2009a}
\begin{equation}\label{eq=2}
    I(\zeta)=I_{\,0}(\zeta)\cdot{{ {\displaystyle \sin^{2}{{N\zeta}\over{2}}} }\over{{\displaystyle \sin^{2}{{\zeta}\over{2}}} }}
\end{equation}
 Function $I_{\,0}(\zeta)$ is an envelope describing diffraction on a single slit and
$$
    \zeta = {{2\pi}\over{\lambda}}\,d\sin(\theta)
$$
 is a phase shift of the waves emitted from two nearest slits, see Fig.~\ref{fig=2}. Observe that Eq.~(\ref{eq=2}) is quite common formula. For example, it describes also revolution of spin of neutrons flying through an $N$-periodic magnetic field~\cite{Agamalyan:1988}.

 Most striking observation in Fig. 1 is existence of fractal structures, sizes of which are smaller than distance between slits.
 As was mentioned above, emergence of the fractal structures have been studied by Berry {\it et al.}~\cite{Berry:2001},
 Amanatidis {\it et al.}~\cite{AmanatidisEtAl:2003}, Sanz~\cite{Sanz:2005}. Sanz, in particular, has devoted the studies whether one can depict by the Bohmian trajectories  the quantum fractal structures. The authors note that universal fractal features of quantum theory
 might be useful in the field of quantum information, for creating efficient quantum algorithms.

As was mentioned, interference effects in the near-field regions can be important with the point of view of quantum computing perspective. Theoretical study of the near-field interference by means of preparation of quantum mechanical models is the first step for an understanding of the quantum computing perspective. Gaussian wavepacket~\cite{Sanz:2007, Sanz:2008} is a simplest model for studying $N$-slit grating interference. Sec.~\ref{sec:level2} deals with the Gaussian wavepacket and its Fourier transforms that give rise to emergence of complex, time-dependent, variance. In Sec.~\ref{sec:level3} we simulate interference by radiation of the Gaussian wavepackets from the $N$-slit grating. Their dispersion in the near-field region produces an interference pattern that manifests itself in a fractal organization of probability density of the wave function. As the detector is shifted in the far-field region the interference pattern transforms to diffraction pattern described by Eq.~(\ref{eq=2}). In that region fractalality disappears. Instead of this, principal maxima of radiation come into being. And they are partitioned by ($N-2$) subsidiary maxima.  Next, the problem is to understand, how does a particle pass through the $N$-slit grating up to a detector. David Bohm had revealed that the particle travels along a single optimal trajectory~\cite{Bohm:1952a, Bohm:1993}, which is named in literature the bohmian trajectory. This approach together with calculation of trajectories is discussed in Sec.~\ref{sec:level4}. Emergence and development of the fractal Talbot patterns are considered in Sec.~\ref{sec:level5} at the slit grating containing many slits and at varying the period $d$ of the grating.
Comparison of classical trajectories and bohmian is given in Sec.~\ref{sec:level6}, concluding section.  In this section we discuss the bohmian trajectories passing through a slit up to the detector, and waviness of the bohmian trajectories caused by exchange of virtual particles with vacuum that is tuned by de Broglie pilot-wave.
\begin{figure}[htb!]
  \centering
  \begin{picture}(100,125)(100,135)
      \includegraphics[scale=0.5]{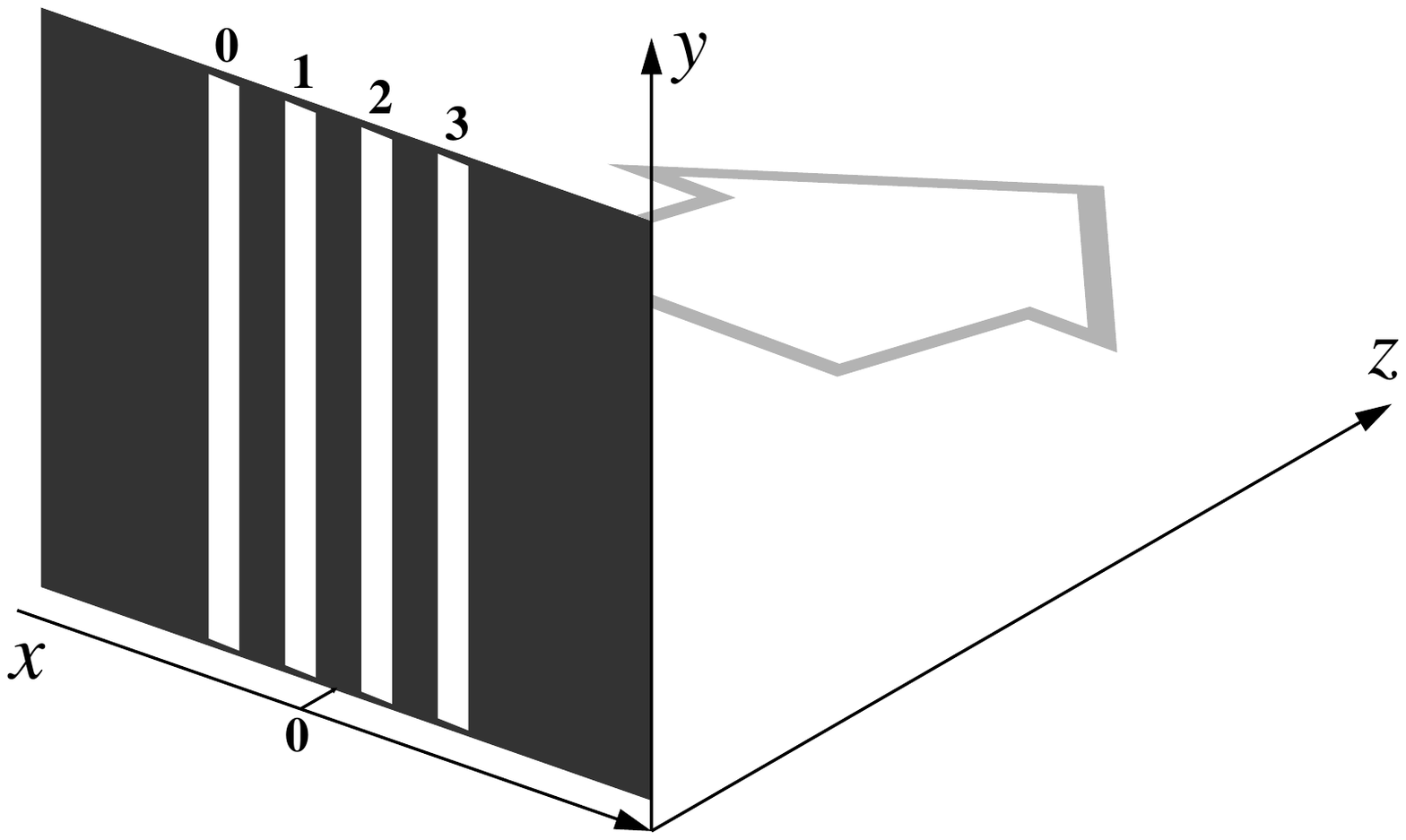}
  \end{picture}
  \caption{
Interference experiment in a cylindrical geometry -  the grating is placed in the plane $(x,y)$ and slits go along the axis $y$. Gray arrow points direction of wave radiation.
 }
  \label{fig=3}
\end{figure}

\section{\label{sec:level2}Gaussian wavepacket and its Fourier transforms.}

As seen in Fig.~\ref{fig=2}, Eq.~(\ref{eq=2}) results from computation of interference of cylindrical waves divergent from the slits. The waves in the vicinity of the slits are no cylindrical, in general. They have a complex form when transforming from a plane wave to cylindrical one. For the sake of simplicity, we suppose that slit borders are fuzzy under Gauss distribution, and a wave from the slit is approximated by the Gaussian wavepacket~\cite{Sanz:2007, Mark:1997}
\begin{equation}\label{eq=3}
    \varphi_{\,0}(x)={\root 4\, \of {{1}\over{2\pi\sigma^{2}}} }
    \exp\Biggl\{ -{{(x-x_{0})^{2}}\over{4\sigma^{2}}} \Biggr\}.
\end{equation}
 Observe that $\varphi_{\,0}^{\,2}(x)=p(x)$ is the probability density distribution that has a mean $x_{0}$ and variance $\sigma^{2}\ge0$.

 We believe that in the immediate neighborhood on each slit, see Fig.~\ref{fig=3}, a wave field is described by the Gaussian wavepacket~(\ref{eq=3}).
 It rely on assumption that edges of the slits are not ideal, but rather fuzzy.

 Observe that the wavepacket~(\ref{eq=3}) disperses as it moves away from the slit to a region pointed out by grey arrow in Fig.~\ref{fig=3}. Let us express this wavepacket by superposition of harmonic waves $\exp\{-2\pi {\bf i} kx_{0}\}$ with wave number $k$ ranging from $-\infty$ to $\infty$. To that end, we execute the Fourier transform
\begin{widetext}
\begin{eqnarray}
\nonumber
  F_{x}\Bigl[\varphi_{\,0}(x)\Bigr](k) &=&
  {\root 4\, \of {{1}\over{2\pi\sigma^{2}}} }\int\limits_{-\infty}^{\infty}
    \exp\Biggl\{ -{{(x-x_{0})^{2}}\over{4\sigma^{2}}} \Biggr\}\exp\{-2\pi {\bf i} kx\}dx \\
    &=& {\root 4\, \of {8\pi\sigma^{2}}}\exp\{-\pi^{2}4\sigma^{2}k^{2}\} \exp\{-2\pi {\bf i} kx_{0}\} = \Phi_{0}(k).
\label{eq=4}
\end{eqnarray}
\end{widetext}
 The function $\Phi_{0}(k)$ is seen to be a harmonic wave $\exp\{-2\pi{\bf i} k·x_{0}\}$ having an amplitude
\begin{equation}\label{eq=5}
    A(k) = {\root 4\, \of {8\pi\sigma^{2}}}\exp\{-\pi^{2}4\sigma^{2}k^{2}\}.
\end{equation}

As we move off from the slit, the wave components $\Phi_{0}(k)$ spread along axis $x$ differently for different $k$.
Let the dispersed wave component be $\Phi_{0}(k)\exp\{- {\bf i}2\pi k\cdot\pi x\}$ . Additional factor $\pi$ scales the dispersion along axis $x$. Observe that $x = vt$ and a speed of the shift along axis $x$ is $v = \hbar k/m$. Here $\hbar$ is the reduced Planck constant, $m$ is a particle mass. As far as $2\pi k\cdot k = (2\pi k)^{2}/2\pi$, a shifted wave component is
\begin{equation}\label{eq=6}
    \Phi_{0}(k)\exp\Biggl\{-{\bf i}(2\pi k)^{2}{{\hbar}\over{2m}}\cdot t\Biggr\}.
\end{equation}

Let us now execute the inverse Fourier transform of the function~(\ref{eq=6})
\begin{widetext}
\begin{eqnarray}
\nonumber
  &F_{k}&\Biggl[\Phi_{0}(k)\exp\Biggl\{-{\bf i}(2\pi k)^{2}{{\hbar}\over{2m}}\cdot t\Biggr\}\Biggr](x) \\
\nonumber
  &=&  {\root 4\, \of {8\pi\sigma^{2}}}
  \int\limits_{-\infty}^{\infty}\exp\{-\pi^{2}4\sigma^{2}k^{2}\}
  \exp\Biggl\{-{\bf i}(2\pi k)^{2}{{\hbar}\over{2m}}\cdot t\Biggr\}
  \exp\{2\pi {\bf i} k(x-x_{0})\}dk \\
   &=& {\root 4\, \of {{2}\over{4\pi\sigma_{\,t}^{2}}} }
    \exp\Biggl\{ -{{(x-x_{0})^{2}}\over{4\,\sigma\sigma_{\,t}}} \Biggr\} = \phi(x,x_{0},t).
\label{eq=7}
\end{eqnarray}
\end{widetext}
 Here
\begin{equation}\label{eq=8}
    \sigma_{\,t} = \sigma\Biggl(1 + {\bf i}\,{{\hbar}\over{2m}}\,t\cdot\sigma^{-2} \Biggr)
\end{equation}
 is a complex time-dependent spreading. It should be noted, that it is a complex variable of time $t$.
 Getting ahead, we can say that this complex variable~\cite{Sanz:2007, Sanz:2008} determines fractal pattern in the near-field region.
 Observe that at $t = 0$ we have  $\sigma_{t=0} =\sigma$ and the function (\ref{eq=7}) comes to   $\phi_{0}(x)$.

\section{\label{sec:level3}Radiation from $N$-slit grating.}

The function~(\ref{eq=7}) is not yet a wave function since it contains no a term describing its translation forward the region, as shown by grey arrow in Fig.~\ref{fig=3}. The term describing such a translation along axis $z$ is represented by a factor $\exp\{{\bf i}\omega t - {\bf i}k_{z}z\}$.
Here $E = \hbar\omega$ is a particle energy and $p_{z} = \hbar k_{z}$ is its momentum along axis $z$. The wave function, in such a case, reads
\begin{eqnarray}\label{eq=9}
  &&\Psi(x,x_{0},z) = \phi(x,x_{0},t)\exp\{{\bf i}\omega t - {\bf i}k_{z}z\} \hspace{24pt} \\
\nonumber  \\
\nonumber
   &=& {\root 4\, \of {{2}\over{4\pi\sigma_{\,t}^{2}}} }
  \exp\Biggl\{ -{{(x-x_{0})^{2}}\over{4\,\sigma\sigma_{\,t}}}\Biggr\}\exp\{{\bf i}\omega t - {\bf i}k_{z}z\}.
\end{eqnarray}
A velocity along axis $z$ is $v_{z} = p_{z}/m = \hbar k_{z}/m$. So, we can express time $t$ in the function~(\ref{eq=9}) via variable $z/v_{z}$.
Therefore, arguments in the function $\Psi(x, x_{0},z)$ do not contain $t$.
\begin{figure}[htb!]
  \centering
  \begin{picture}(100,185)(85,97)
      \includegraphics[scale=0.45]{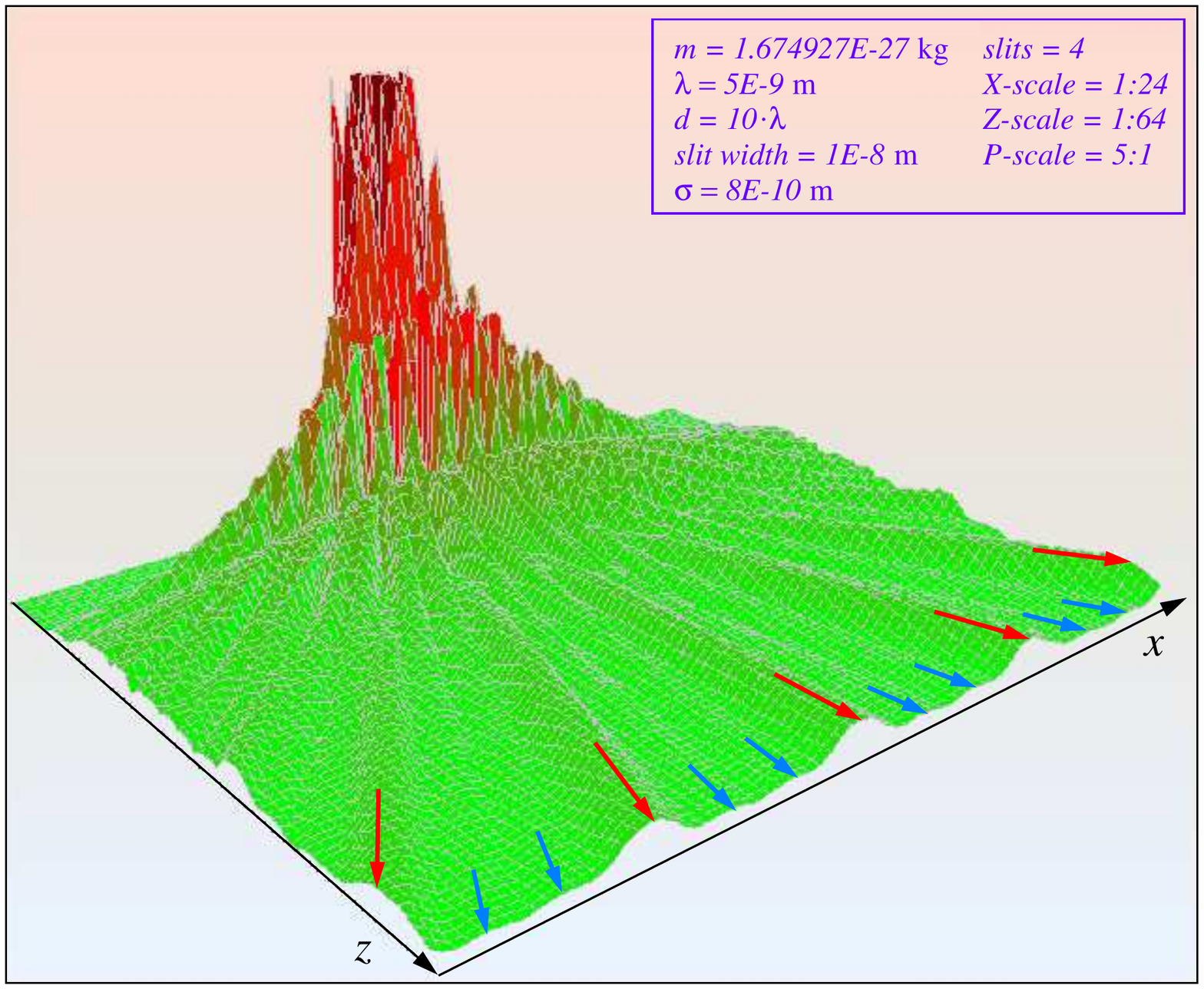}
  \end{picture}
  \caption{
Diffraction pattern in a transient region, $(x;z)\in  ((0\cdots1.4\times 10^{-6});(5\times 10^{-8}\cdots3.8\times 10^{-6}))$ m,
from near-field to far-field ones. Red arrows point out to directions of radiation along the principal peaks. Blue arrows point out  to directions of weak radiation along the subsidiary maxima.  }
  \label{fig=4}
\end{figure}

Putting the frame of axis $x$ in center of the slit grating, containing $(N+1)$ slits, ($n = 0,1,2,\cdots,N$), as shown in Fig.~\ref{fig=3}, we find that position of $n^{\rm th}$ slit is $x_{0}=(n - N/2)d$. Here distance $d$ is period of the grating.
Superposition of the waves~(\ref{eq=9}) emitted by all $(N+1)$ slits reads
\begin{equation}\label{eq=10}
    |\Psi(x,z)\rangle = {{1}\over{N+1}}\sum\limits_{n=0}^{N}
 \Psi\Biggl(x,\Biggl(n-{{N}\over{2}}\Biggr)d,z\Biggr).
\end{equation}
 Probability density
\begin{equation}\label{eq=11}
    p(x,z)=\langle \Psi(x,z)|\Psi(x,z)\rangle
\end{equation}
 is an observable. Fig.~\ref{fig=4} shows the probability density $p(x,z)$ which is calculated within a transient region from the near-field region to the far-field ones.
The grating contains four slits. It should be noted, that coordinate $z$ in this figure begins from a value that is slightly more than zero. The whole point is that the wave function at $z = 0$ has singularities located on the slits.

 Here we have simulated neutron scattering on the slits, with the neutrons having a wavelength $\lambda=5 {\rm nm}$. Kinetic energy of the neutrons is about $5.25\times 10^{-24}$ J.  It means, that the neutrons are very cold
 \footnote[2]{http://en.wikipedia.org/wiki/Thermal{\_}neutron },
 i.e., temperature of the neutrons is about $T = E / kB \approx 0.38$ K. Here $kB   \approx 1.38\times10^{-23}$ JK$^{-1}$ is the Boltzmann constant.
\begin{figure}[htb!]
  \centering
  \begin{picture}(100,70)(123,220)
      \includegraphics[scale=0.6]{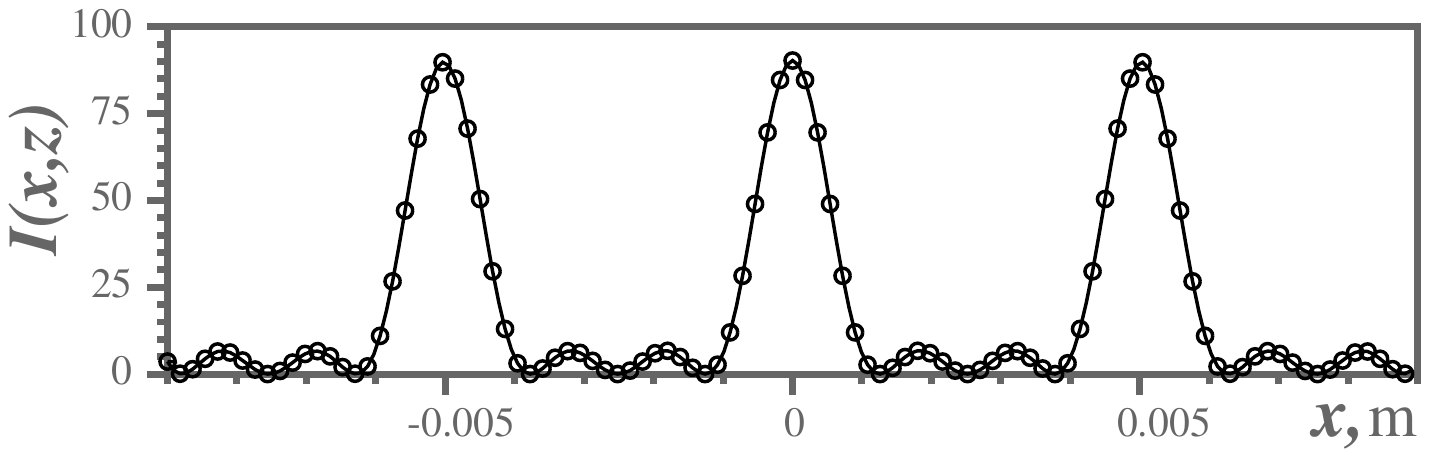}
  \end{picture}
  \caption{
Diffraction in the far-field region, $z = 0.004$ m. Cross-section of the diffraction pattern (see previous figure)  is displayed by circles. Solid curve shows the same cross-section calculated by the diffraction formula~(\ref{eq=2})
with parameters $\zeta$   and $I_{0}$ defined in Eqs.~(\ref{eq=12}) and~(\ref{eq=13}).
  }
  \label{fig=5}
\end{figure}

 Fig.~\ref{fig=4} demonstrates diffraction pattern in a transition region from the near-field region to the far-field one. Red arrows point out to the principal maxima. And blue arrows point out to the subsidiary maxima. Alternation of the principal maxima and subsidiary ones is described by
 Eq.~(\ref{eq=2}). Parameters $\zeta$  and $I_{0}$, in this case, have the following forms~\cite{Sbitnev:2009a}
 \begin{equation}\label{eq=12}
    \zeta(x,z)=xd\,{{m\hbar (z/v_{z})}\over{D(z)}}
 \end{equation}
 and
 \begin{equation}\label{eq=13}
    I_{0}(x,z)=\sqrt{{{1}\over{\pi}}} {{m\sigma}\over{\sqrt{D(z)}}} \exp\Biggl\{-{{2m^{2}\sigma^{2}x^{2}}\over{D(z)}}\Biggr\},
 \end{equation}
 where denominator $D(z)$ is as follows
 \begin{equation}\label{eq=14}
    D(z) = m^{2}\sigma^{4}+\hbar^{2}(z/v_{z})^{2}/4.
 \end{equation}
 Here $(z/v_{z})=t$ is a flight time along the path length $z$.
 Diffraction curve~(\ref{eq=2}), at substituting terms $\zeta$ and $I_{0}$ by functions~(\ref{eq=12})-(\ref{eq=13}), has been calculated in the far-field region,  $z = 0.004$~m, is shown in Fig.~\ref{fig=5}.
 Circles in this figure relate to results simulated by means of the Gaussian wavepacket.
\begin{figure}[htb!]
  \centering
  \begin{picture}(100,185)(85,97)
      \includegraphics[scale=0.45]{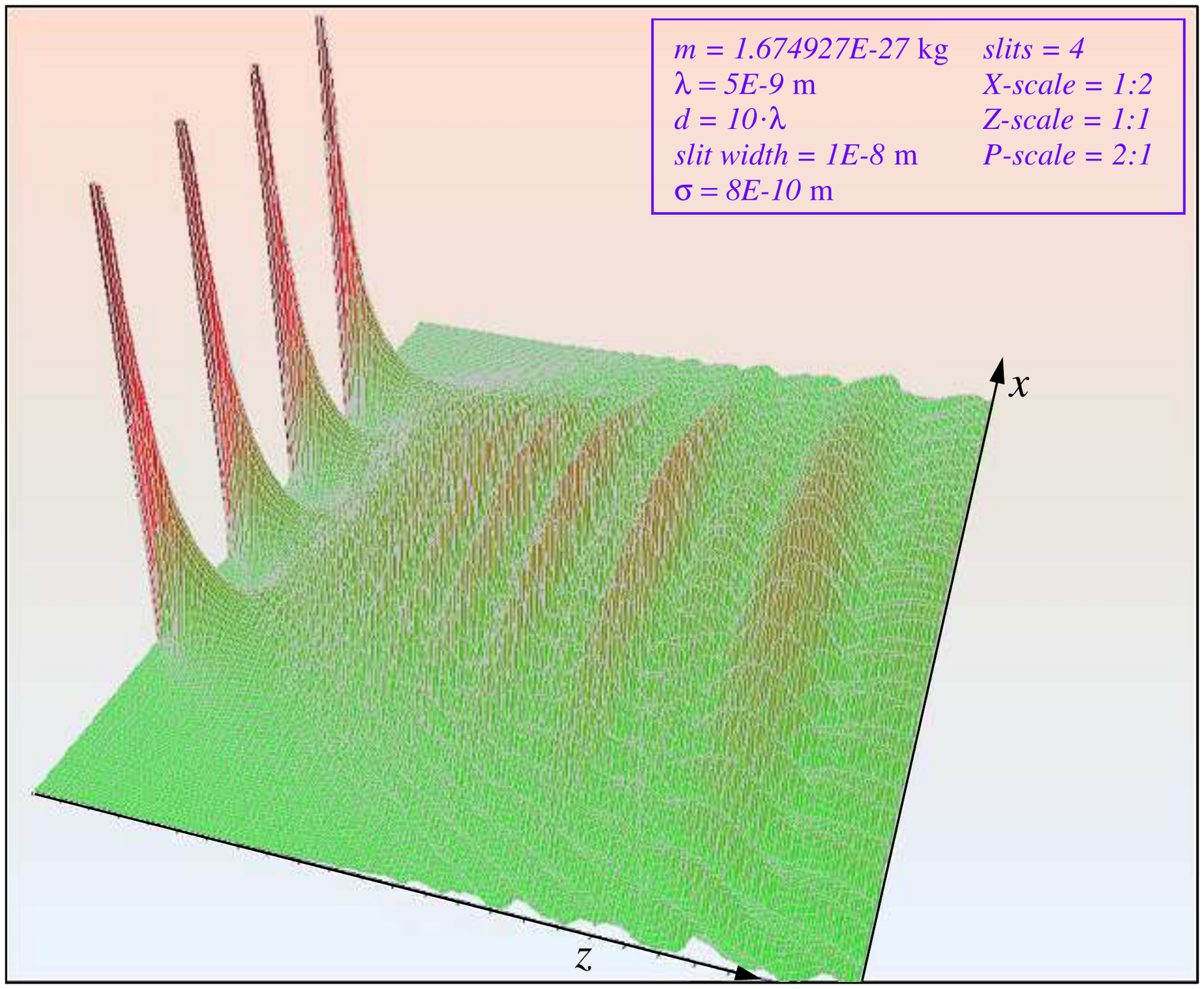}
  \end{picture}
  \caption{
Diffraction pattern in the near-field region, $(x,z)\in ((0\cdots3\times 10^{-7}) ;( 8\times 10^{-10}\cdots2\times 10^{-7}))$ m. Ripples of the probability density lengthen as distance to the N-slit sources is increased.
  }
  \label{fig=6}
\end{figure}

 Turn back to Fig.~\ref{fig=4}. In order to show the diffraction pattern in the far-field region, we have coarsen resolution of the probability density distribution. For this reason, we see a rough pile-up of maxima in the near-field region. Fig.~\ref{fig=6} shows the same pattern at more detailed resolution in the near-field region. One can see, the probability density distribution increases catastrophically nearby the slits. At some distance from the slits, their radiations are superimposed with each other. Such a superposition manifests itself by emergent peaks. The peaks near the slits seem short. And they become longer, as a look moves from the slits to the far-field region. As soon as the detector is shifted to the far-field region, the peaks are transformed to typical diffraction rays going away to infinity.

\section{\label{sec:level4}Density distribution plot and bohmian trajectories.}

Let us project the density distribution $p(x,z)$ to the plane $(x,z)$. Fig. 7 demonstrates this projection in grey palette. Dark places show  maximal values of $p(x,z)$, black patches, in particular, side with the slits. Light places relate to minimal values. Fractality in the interference pattern are well visible when passing from the slit sources to the right edge of the figure.
\begin{figure*}[htb!]
  \centering
      \includegraphics[scale=0.65]{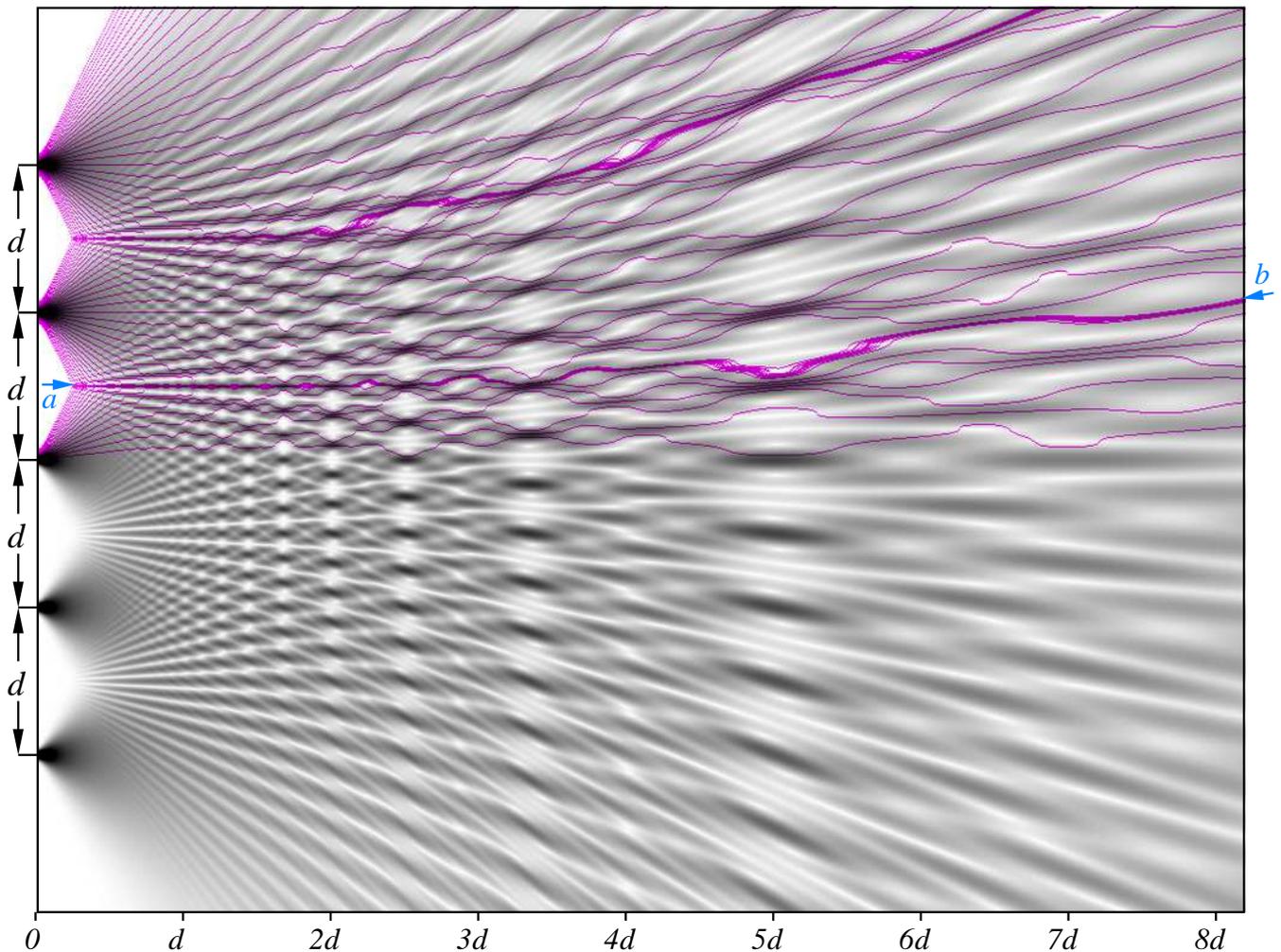}
  \caption{
Density distribution plot drawn by grey palette displays the Talbot effect: white color relates to zero intensity, and black color points to maximal intensity. Violet curves drawn in the upper part of the plot depict the bohmian trajectories.
Distance $d = 5\times10^{-8}$  m. It is equal to $10\cdot\lambda$, $\lambda=5\times10^{-9}$ m..
  }
  \label{fig=7}
\end{figure*}

Violet curves traced in the upper part of the figure depict bohmian trajectories.
Finding the trajectories is based on variation of the action integral~\cite{Sbitnev:2009b}, which leads to the principle of least action.
A general formula computing the bohmian trajectory~\cite{Sanz:2008, Davidovich:2008},  the guidance equation~\cite{Valentini:2009, StruyveAndValentini:2009}, reads:
\begin{equation}\label{eq=15}
    v_{x}={\dot{x}}={{\hbar}\over{m}}\Im \biggl( |\Psi(x,z)\rangle^{-1}\nabla |\Psi(x,z)\rangle \biggr)
\end{equation}
 As a result, we can find a current position of the particle by the following formulas:
\begin{equation}\label{eq=16}
\left\{
  \begin{array}{c}
    x(t) = x_{0} + {\displaystyle \int\limits_{0}^{t}} v_{x} d\tau, \\ \\
    z(t) = z_{0} + v_{z}\cdot t. \hspace{12pt} \\
  \end{array}
\right.
\end{equation}
 The velocity along $z$ is $v_{z} = \hbar k_{z}/m$.
 It stems from the term $\exp{{\bf i}\omega t - {\bf i}k_{z}z}$ as adopted before, see Eq.~(\ref{eq=9}).
\begin{figure*}[htb!]
  \centering
      \includegraphics[scale=0.70]{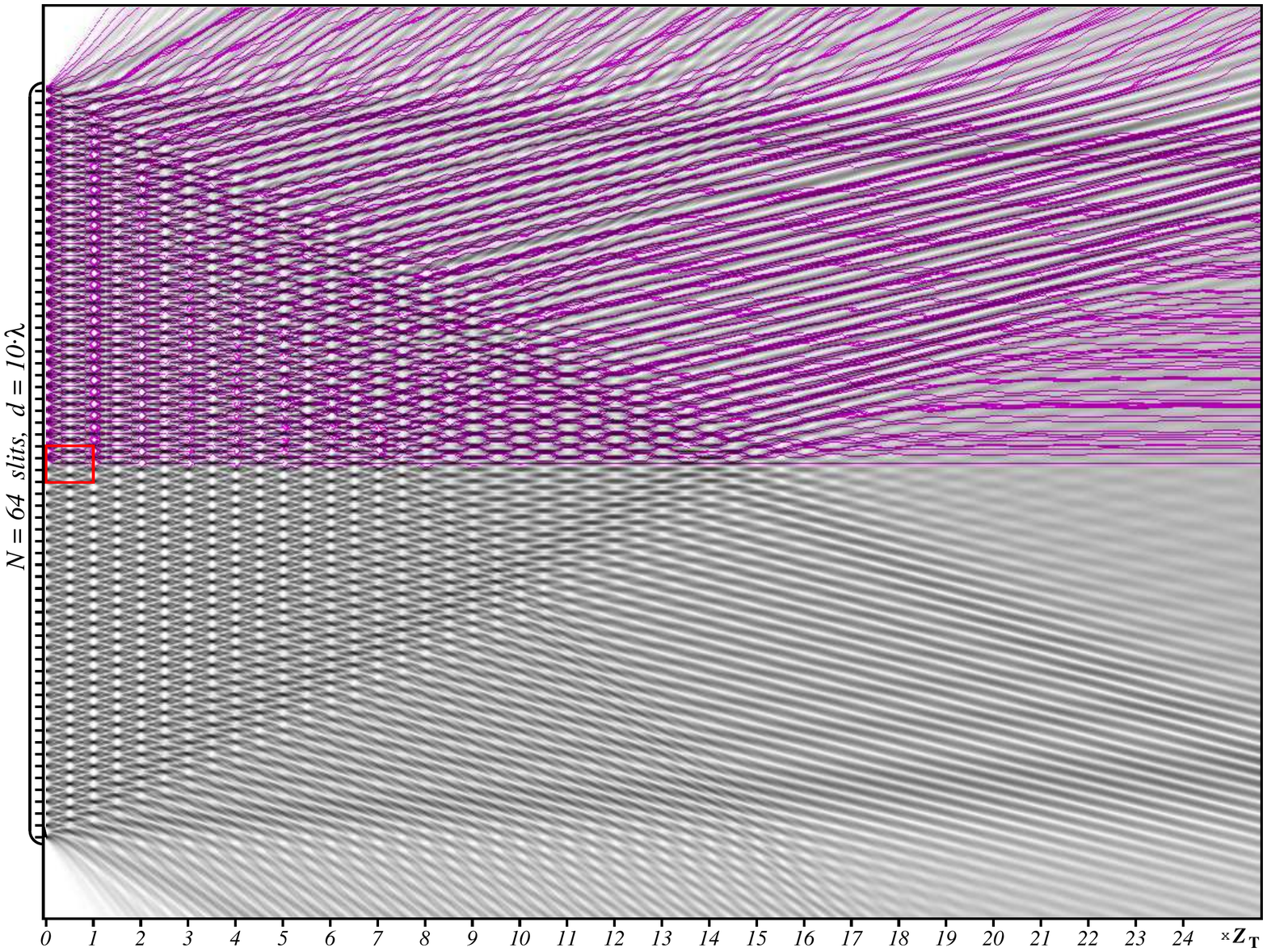}
  \caption{
Interference from $N=64$ slits in the near-field region.
Density distribution plot drawn by grey palette displays the Talbot effect: white color relates to zero intensity, and black color points to maximal intensity. Violet curves drawn in the upper part of the plot depict the bohmian trajectories.
Grating parameters are $\lambda=5$ nm, $d = 10\lambda = 50$ nm, $z_{\rm T}=1000$ nm.
  }
  \label{fig=8}
\end{figure*}

 Velocity $v_{x}$ is seen from Eq.~(\ref{eq=15})  to be (a) proportional to gradient of the wave function; and (b) inversely proportional to the same wave function. It means: (a) a trajectory undergoes greatest variations in parts, where the wave function demonstrates slopes; and (b) the trajectory avoids areas, where the wave function tends to zero. Violet wavy curves in Fig.~\ref{fig=7} manifest themselves the above properties well enough. One can see, the wavy curves group predominantly in dark-grey spots and avoid white spots. In extreme cases, the trajectories traverse the white spots almost transversally.

 Does this bohmian trajectory pattern relate to the many-worlds theory? Answer is negative~\cite{Valentini:2009, Kent:2009}. For that aim, let us scatter particles on the N-slit grating by single-piece. Let a single particle pass, for example, the central slit nearby the top border. As is shown in Fig.~\ref{fig=7}, the particle scatters to a direction pointed out by blue arrow {\it a}. Next, the particle goes on a wavy stream and follows to a place pointed out by blue arrow {\it b}. Here we suppose that, movement of the particle submits to the principle of least action. Therefore, a number of problems arises right now. The first problem arising here is as follows: what is cause which forces the particle to change its own direction in the vicinity of point {\it a}? And the second problem is: what is cause which forces the particle to perform wavy motions? We confirm that, any experimental setup for a quantum mechanical experiment is, in fact, the quantum instrument. The setup contains, in our case, $N$-slit grating with its parameters fitted for a wavelength of the particle.  Boundary conditions of the setup determine the wave function on the edges. In whole, the experimental setup determines configuration of the wave function given in the working space. The function is named de Broglie pilot-wave~\cite{Valentini:2009, StruyveAndValentini:2009}.  In fact, its squared image, the density distribution~(\ref{eq=11}), is shown in Fig.~\ref{fig=7} in grey palette. So, the bohmian trajectory shows an optimal path for a particle, which is guided by the de Broglie pilot-wave within the working space.

\section{\label{sec:level5}Fractals in the near-field region.}

 In order to study fractal patterns arising in the near-field region in detail, amount of slits in the grating is not enough.
\begin{figure}[htb!]
  \centering
  \begin{picture}(100,70)(125,250)
      \includegraphics[scale=0.66]{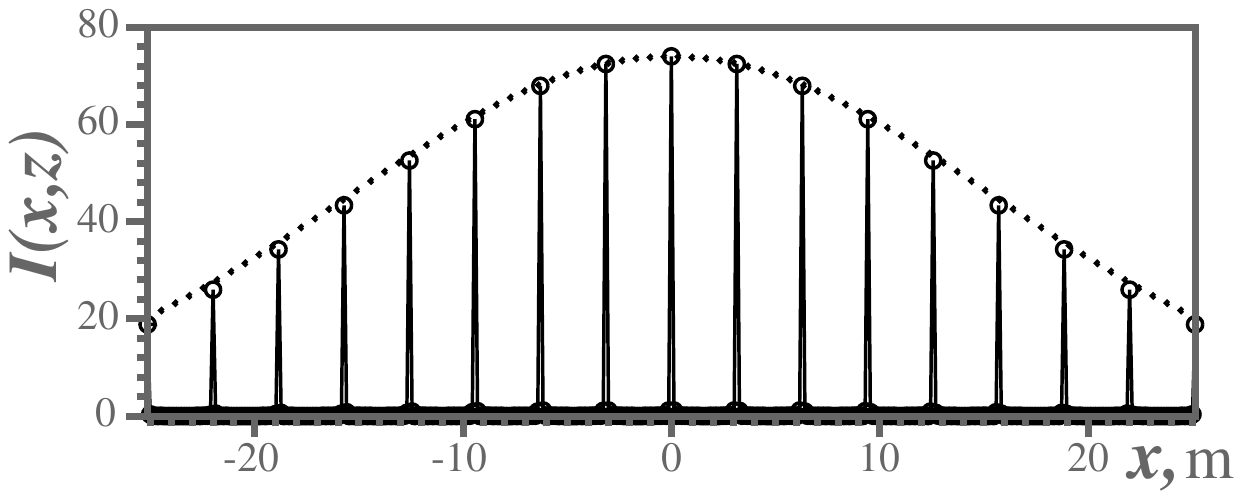}
  \end{picture}
  \caption{
 Diffraction in the far-field region, $z = 1.25$ m. Cross-section of the diffraction pattern is displayed by circles.
 Solid sharp principal peaks show the same cross-section calculated by the diffraction formula~(\ref{eq=2})
 with parameters $\zeta$   and $I_{0}$ defined in Eqs.~(\ref{eq=12}) and~(\ref{eq=13}).
 Dot envelope curve is $I_{0}(x)\cdot N^{\,2}$.
  }
  \label{fig=9}
\end{figure}
 In fact, the amount should tend to infinity. We will consider here, however, emission from a finite grating containing 64 slits, see~Fig.~\ref{fig=8}. One can see, in the vicinity of the slits placed in the middle of the grating (in the figure that place is drawn by red square) the Talbot carpets can be  perfect enough. As the wave front spreads to an area reaching to the far-field region, the Talbot ordered structure is dissolved by triangle-like manner, as is described in~\cite{Sanz:2007}.
 In the far-field region we will observe an usual diffraction from $N$-slits, i.e., a set of principal maxima partitioned from each other by $(N-2)$ subsidiary maxima, Fig.~\ref{fig=9}.

 Violet wavy curves drawn in the upper part in Fig.~\ref{fig=8} are Bohmian trajectories. They occupy preferably regions, where the probability density reaches local maxima. And vice-versa, they avoid local minima. Instructive to compare behavior of these Bohmian trajectories with those shown in Sanz's and Mil{\'e}t's article~\cite{Sanz:2007}.

\subsection{\label{subsec:level5.1}Talbot carpets.}
\begin{figure}[htb!]
  \centering
  \begin{picture}(100,170)(90,20)
      \includegraphics[scale=0.5]{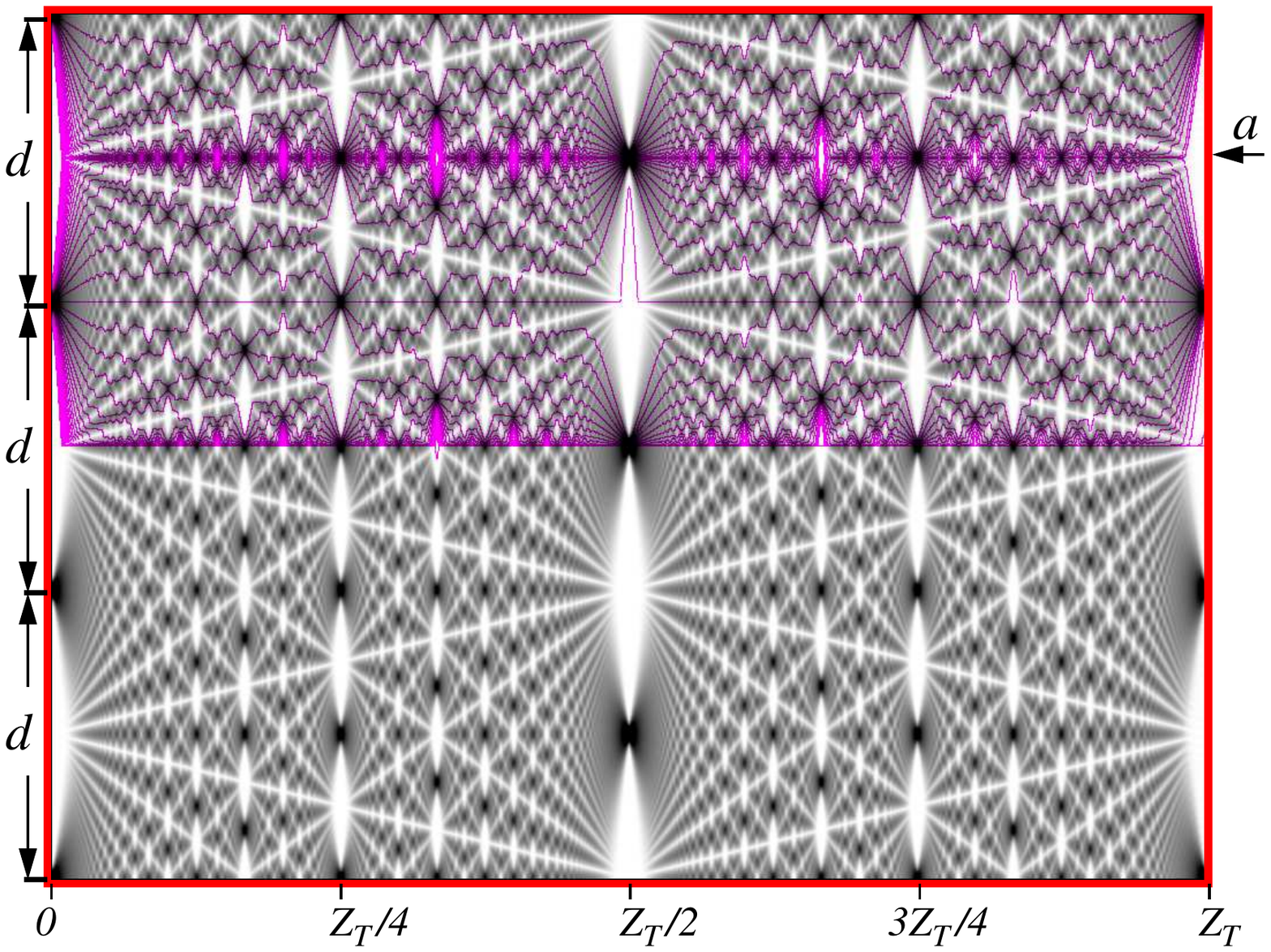}
  \end{picture}
  \caption{
  Talbot carpet:
  $d=10\lambda=50$~nm, $z_{\rm T}=1000$~nm.
 }
  \label{fig=10}
\end{figure}
\begin{figure}[htb!]
  \centering
  \begin{picture}(100,170)(90,20)
      \includegraphics[scale=0.5]{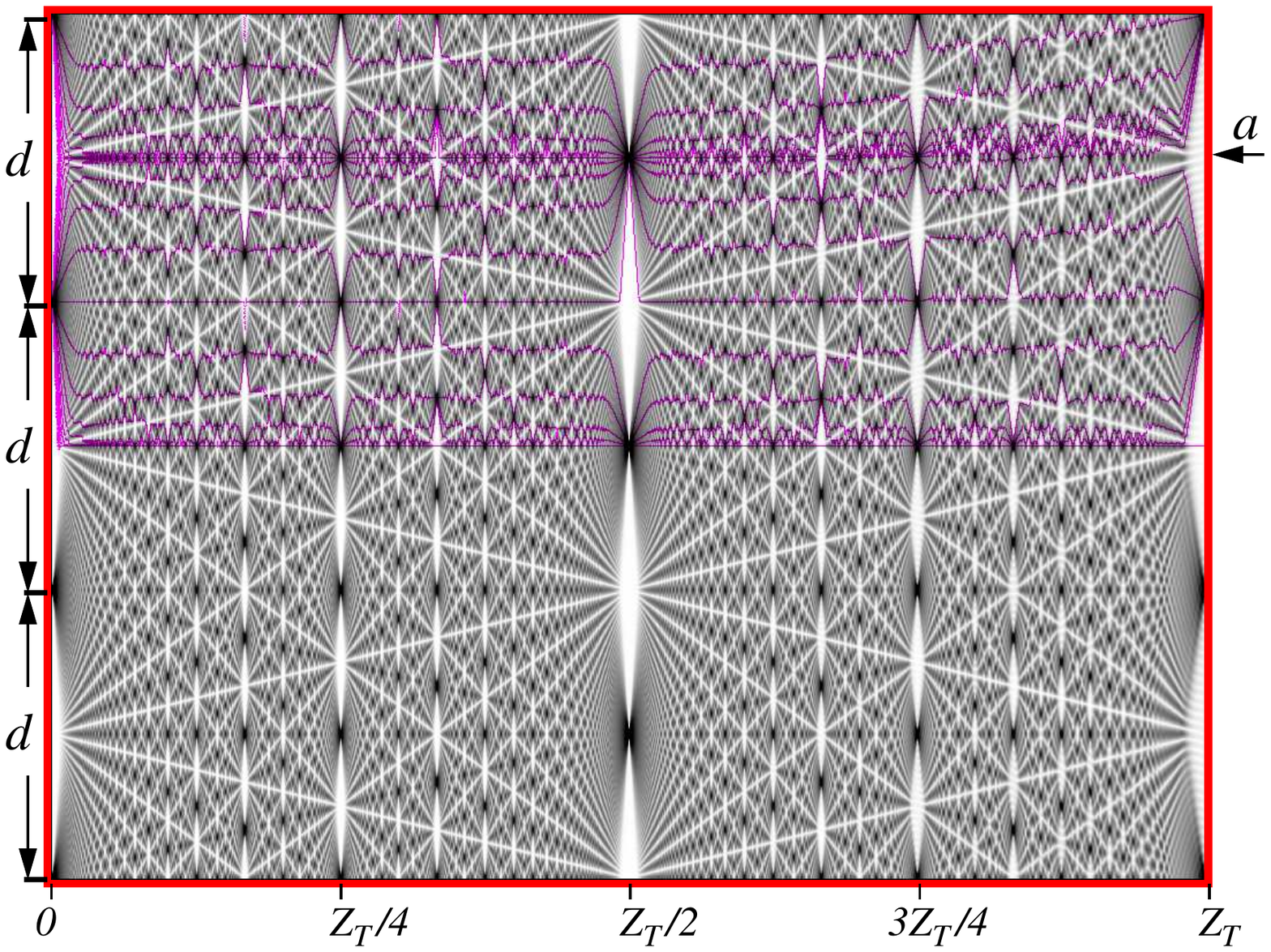}
  \end{picture}
  \caption{
  Talbot carpet:
  $d=20\lambda=100$ nm,
  $z_{\rm T}=4000$ nm.
 }
  \label{fig=11}
\end{figure}
\begin{figure}[htb!]
  \centering
  \begin{picture}(100,170)(90,20)
      \includegraphics[scale=0.5]{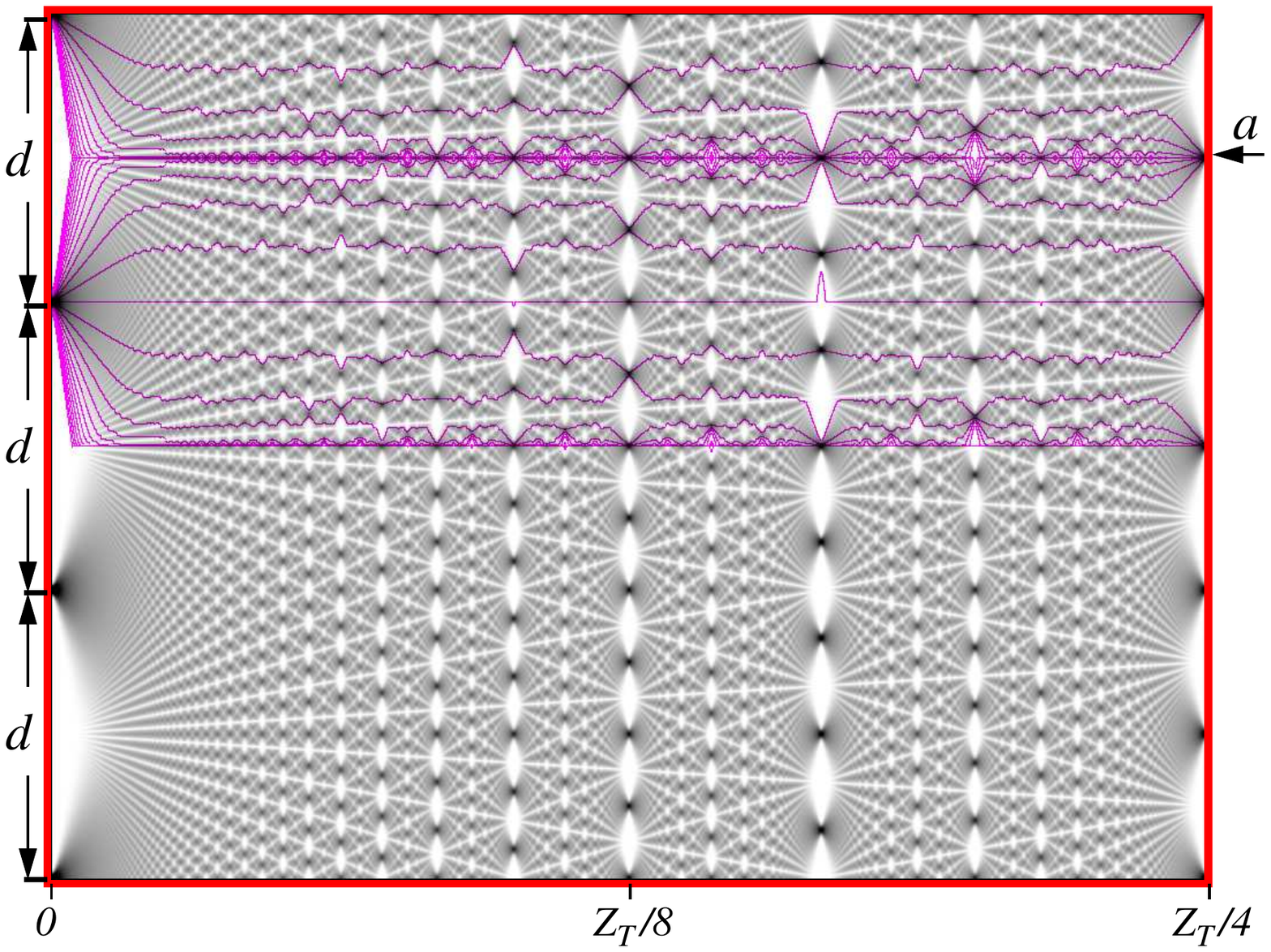}
  \end{picture}
  \caption{
  Talbot carpet:
  $d=40\lambda=200$ nm,
  $z_{\rm T}=16000$ nm.
 }
  \label{fig=12}
\end{figure}
\begin{figure}[htb!]
  \centering
  \begin{picture}(100,220)(75,10)
      \includegraphics[scale=0.85]{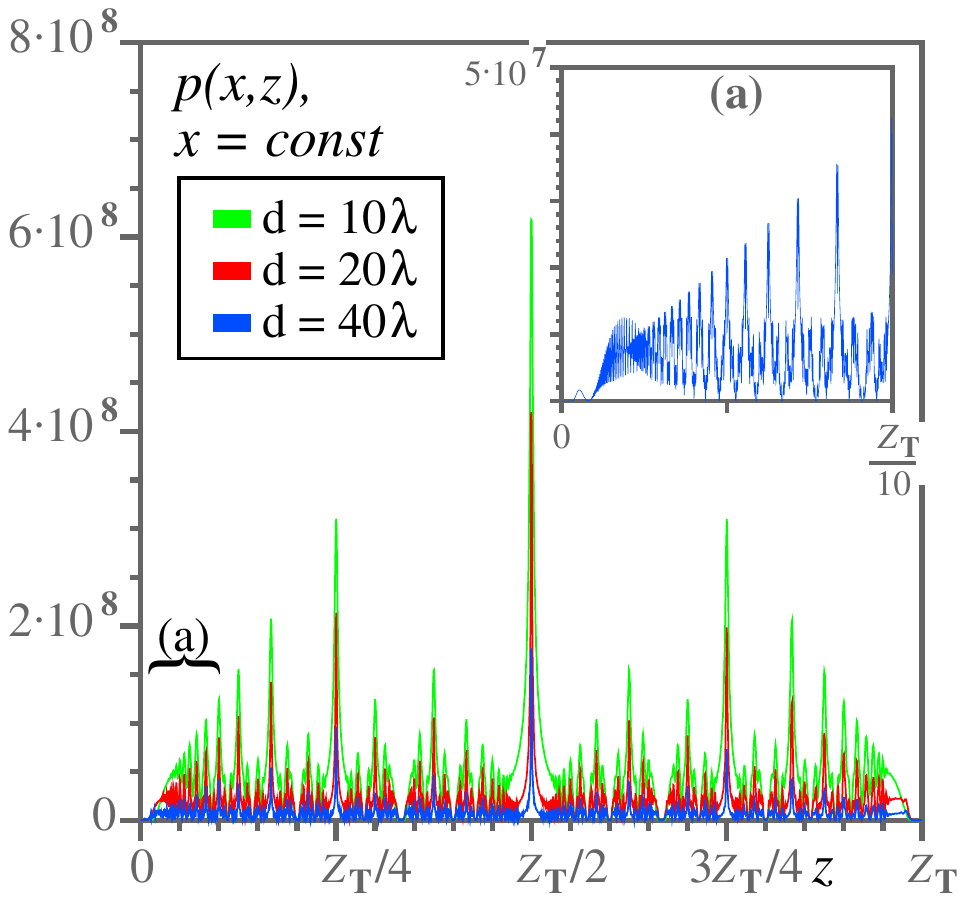} 
  \end{picture}
  \caption{
  Probability density distributions along cross-section pointed out by arrow $a$ in Figs.~\ref{fig=10},~\ref{fig=11}, and~\ref{fig=12}
  are shown by green, red, and blue colors, respectively.
  Insert (a) shows behavior of $p(x,z)$ near the origin of coordinates.
  }
  \label{fig=13}
\end{figure}
\begin{figure*}[htb!]
  \centering
  \begin{picture}(100,400)(200,10)
      \includegraphics[scale=0.85]{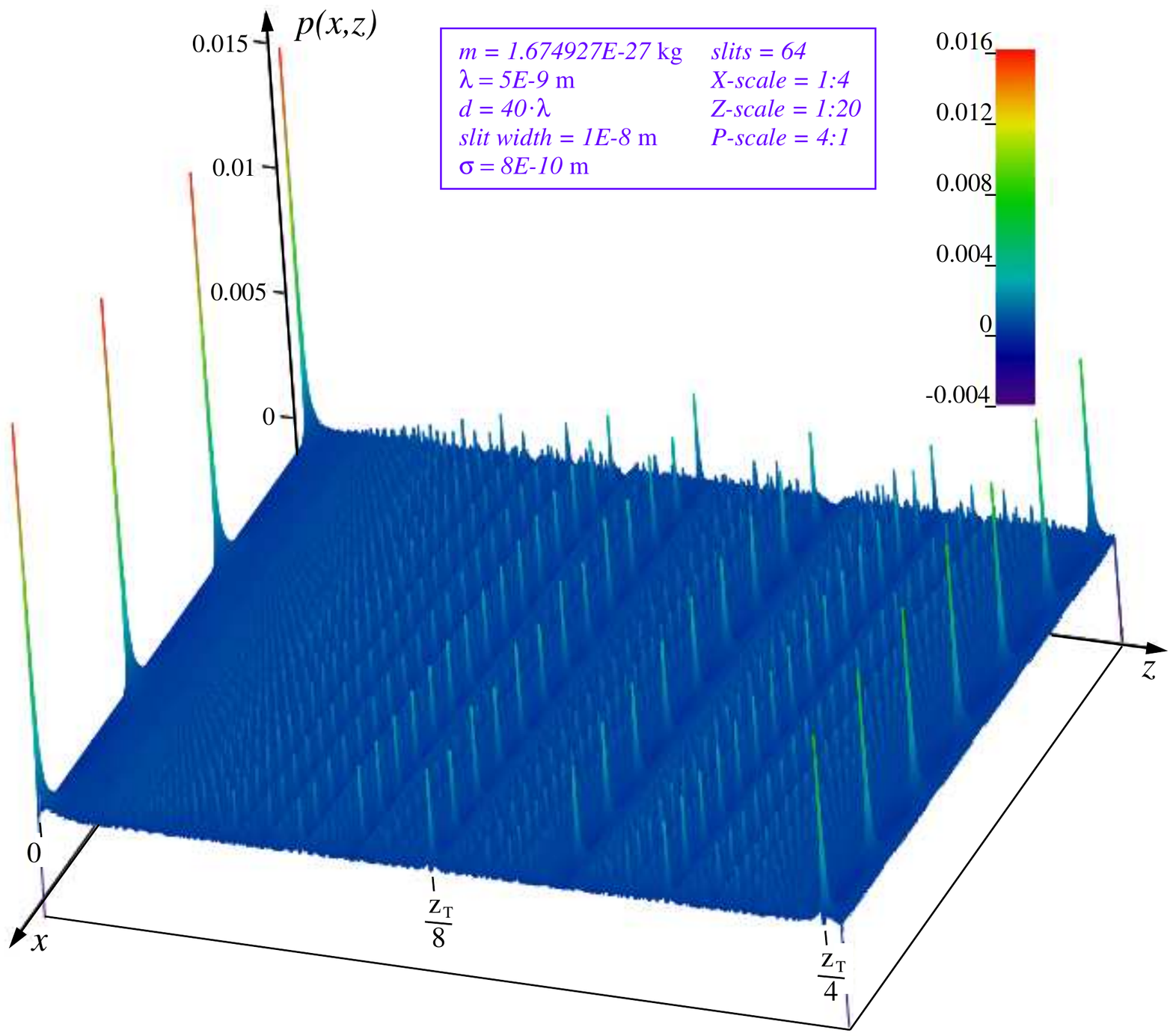}
  \end{picture}
  \caption{
  Probability density distribution  
  approaches infinite set of $\delta$-functions as $\lambda/d$ tends to zero.
  }
  \label{fig=14}
\end{figure*}

 Let us now discuss the Talbot patterns cut from the red square adjoining to the slit screen, see Fig.~\ref{fig=8}. Here we study emergence of the Talbot carpets at different relations of the period $d$ and wavelength $\lambda$. Since the slit grating contains only $N=64$ slits, we can study a small set of such relations, $d=k\lambda$, $k$ is integer. We have given $d=10\lambda$, $d=20\lambda$, and $d=40\lambda$. According to Eq.~(\ref{eq=1}), we have $z_{\,\rm T} = 2k d$. In the Talbot's units we can compare the Talbot carpets calculated for different input parameters.
 Figs.~\ref{fig=10},~\ref{fig=11}, and~\ref{fig=12} demonstrate the Talbot carpets calculated at $\lambda=5$~nm and $d=10\lambda=50$~nm, $d=20\lambda=100$~nm, $d=40\lambda=200$~nm, respectively.

 Instructive to compare Talbot carpets shown in Figs.~\ref{fig=1} and~\ref{fig=10}.
 The both figures are seen to relate to each other as negative and positive images.
 One can see, the both carpets show equivalent details. Whereas, the Talbot carpet shown in Fig.~\ref{fig=11} is seen to contain more subtle details. In other words, this carpet is more fine-grained, than previous one.
 The Talbot carpet shown in Fig.~\ref{fig=12} contains so much subtle details that the figure has been drawn only in the first quarter of the Talbot Length $z_{\,\rm T}$. Here we can see fractal organization of the Talbot carpet very clearly.

 Behind the $N$-slit grating, the field we can see in Figs.~\ref{fig=10},~\ref{fig=11} , and~\ref{fig=12} has a fractal structure of fantastic complexity.
 Let us glance on behavior of the probability density distribution along a cross-section pointed out by arrow $a$ in these Figs. We see, that the probability density vanishes at points $z=0$ and $z=z_{\,\rm T}$ and it reaches maximal values
 at $z=z_{\,\rm T}/4$, $z=z_{\,\rm T}/2$, $z=3z_{\,\rm T}/4$. In the other points the probability density alternates minimal and maximal values by irregular manner. More strictly, the irregularity exposes fractal nature.
 Fig.~\ref{fig=13} shows behavior of the probability density distribution along the mentioned cross-section.
 Insert (a) shows its behavior near the origin of coordinates, $z\in (0, z_{\,\rm T}/10)$.
 One can guess, that the probability density distribution tends to Cantor-like set, as $d/\lambda$ tends to infinity. This is in good agreement with a statement given by Berry {\it et al.} in~\cite{Berry:2001},
 that proclaims that the Talbot fractal emerges as the ratio $\lambda/d$ approaches zero and when the number $N$ of illuminated slits tends to infinity as well~\cite{BerryKlein:1996, Berry:1996}. At $N$ finite, however, the Talbot patterns are blurring and defocusing. And what is more, the Talbot effect is destroyed catastrophically as an observer shifts the detector either to edge of the $N$-slit grating or beyond the near-field region.

 The Cantor set is a set of points lying on a single line segment that has a number of remarkable and deep properties
 \footnote[3]{http://en.wikipedia.org/wiki/Cantor{\_}set}. First, the Cantor set cannot contain any interval of non-zero length. On the other hand, integral of the probability density distribution throughout all physical space has to be equal to unit.
 From here it follows, that the probability density distribution approaches an infinite set of $\delta$-functions, see Fig.~\ref{fig=14}, as the ratio $\lambda/d$ tends to zero.

 Physically, limit of $\lambda\rightarrow 0$ is not available. It should be noted here, that at the wavelength tending to zero, kinetic energy of the incident particles approaches infinity.  In that case, the particle beam will heat up the grating. And secondly, if the kinetic energy is increased further, the particles begin to destroy the grating.

 In the upper parts of Figs.~\ref{fig=10},~\ref{fig=11}, and~\ref{fig=12}, Bohmian trajectories representing particle's paths have been drawn.
 They are pictured by violet dots tracking predominantly along dark places relating to heightened values of the probability density.
 Their behavior is well visible in the vicinity of the dark nodes localized at cross-sections $z_{\,\rm T}/4$, $z_{\,\rm T}/2$, $3z_{\,\rm T}/4$.
 In general, particles jump along separated points until they leave the near-field region. Real possibility is that, the particle is tunneling throughout the suppressed intervals. In other words, it reaches the far-field region along fantastical zigzag paths, that are frequently interrupted by tunneling.
  One can see, behavior of the Bohmian trajectories is complex enough in the fractal media. Nevertheless, it is predictable, since it is based on solution of two coupled equations - the quantum Hamilton-Jacobi equation and the continuity equation. The both result from the Schr{\"o}dinger equation.

\section{\label{sec:level6}Concluding discussion.}

 A fractal, as defined by Mandelbrot, "is a shape made of
 parts similar to the whole in some way"~\cite{Addison:1997}. An exact fractal is an "object
 which appears self-similar under varying degrees of magnification
 ..... in effect, possessing symmetry across scale,
 with each small part replicating the structure of the whole"~\cite{Addison:1997}.
 In this key
 Berry has also written in a private communication the following assertion "Fractality requires three conditions: infinitely many slits, paraxial
 propagation, and discontinuous initial conditions (i.e. sharp slits).
 Then the fractal structure, including fractal dimensions, can be calculated
 explicitly."
 As for the fractal Talbot effect, Berry {\it et al.} have written in~\cite{Berry:2001} "Quantum and optical carpets provide a dramatic illustration
of how limits in physics that seem familiar can in fact be complicated
and subtle. It is no exaggeration to say that perfect
Talbot images, and infinite detail in the Talbot fractals, are
emergent phenomena: they emerge in the paraxial limit as
$\lambda/d$ (here $d$ is signed instead of $a$, V.~S.) approaches zero.
At first this seems paradoxical, because the short-wave
approximation is usually regarded as one in which interference
can be neglected, whereas the Talbot effect depends
entirely on interference. The paradox is dissolved by noting
that the Talbot distance increases as the wavelength approaches
zero, so here we are dealing with the combined limit
of short wavelength and long propagation distance: in the
short-wavelength limit, the Talbot reconstructed images
recede to infinity."

 Talbot patterns arising in the near-field region demonstrate the above mentioned signs fully.
 These patterns have emergent at simulation of scattering cold neutrons on many-slit grating (we have chosen parameters relating to this particle).
 A question arises here - how does a single particle travel behind the grating?

First of all, let us recall some solutions from the classical mechanics~\cite{Lanczos:1970}.

Geodesic trajectories in classical mechanics to be found by applying the principle of least action~\cite{Lanczos:1970} are real trajectories of mechanical objects. A beam of the trajectories, that tracks a relief, stems from the initial conditions slightly differing from each other. Observe that the geodesic trajectories nowhere never intersect.

The number of trajectories, crossing a surface $\delta{S}$, is conserved, no matter how the surface deforms at moving along the beam. In fact, this observation demonstrates the conservation law of number of the trajectories passing through the surface, i.e., the trajectories never disappear and does not appear again. Violation of this law in classical physics would be tantamount to recognize teleportation - a classic body disappear suddenly and unexpectedly announced again.

Interpretation of the continuity equation of the density of trajectories is guided by incompressible fluid, which flows along routes specified by the geodesic trajectories. All physical space we can imagine is filled with this fluid~\cite{Lanczos:1970, Wyatt:2005}.
The basis for such an idealization is experience which shows, that there is a rather broad class of fluids for which even large changes in pressure do not lead to significant change in density. This fluid fills the environment continuously. Its molecular structure, at that, is ignored. In classical physics the continuity equation does not determine the fate of trajectories in the future. Their fate is determined only by the Hamilton-Jacobi equation. Other situation arises in quantum mechanics. Here the both equations, the continuity equation and the Hamilton-Jacobi equation, connected with each other via the bohmian quantum potential~\cite{Bohm:1952a, Bohm:1993}, take part in determination of the geodesic trajectories.
First scientist was Madelung who had derived the same set of equations in 1926~\cite{Madelung:1926}.

In accordance with the de Broglie-Bohm interpretation, the wave function sets up a  $\Psi$-field, which fills physical space of the experimental setup. Its squared modulus, the probability density distribution, is shown in Figs.~\ref{fig=7},\ref{fig=8},\ref{fig=10}--\ref{fig=12}, in gray palette. All physical space we can guess is filled by a fluid-like background with the density distribution $|\Psi|^{2}$~\cite{Madelung:1926, Bohm:1954}. It is important to note, that the state of such a fluid is determined by geometry of the experimental setup. And route of the trajectories depends on the density of the fluid, its gradients, in the neighborhood of each point of the experimental setup. The guidance equation allows to find the trajectories that penetrate the field by the best way. In turn, the density distribution depends on the route of the trajectories. They are mutually dependent.

Such a fluid is seen to be vacuum. Polarization of vacuum is an extraordinary phenomenon since Hendrik Kasimir had proposed the existence of physical forces arising from a quantized field~\cite{Casimir:1948}. From this perspective, change of direction of a trajectory of a particle can be expressed in terms of exchange of virtual particles in vacuum. The vacuum has a vastly rich structure. It, implicitly, has all of the properties that a particle may have. This situation can be expressed by a sentence - the vacuum contains relative '{\it nothing}', and at the same time, the potential '{\it all}'.

The Feynman path integral~\cite{Feynman:1965} is a way to understand many manifestations of the vacuum.  In fact, a proposition is as follows: virtual pairs emergent in a short time and annihilating in the end of the time loop are pairs permitting the particle to test a trajectory both forward and backward in time~\cite{Ord:2003, Ord:2004, WerbosDolmatova2000}. The extra degree of freedom obtained by allowing both directions in time gives the particle to get a non-local information needed for establishing an optimal path from a source to detector without the involvement of intelligent particles, intelligent observers, or multiple universes.

\begin{acknowledgments}
 The author thanks Dr. A. S. Sanz, who impelled more deep study of fractal interference patterns in the near-field region.
 The author thanks also Prof. M. V. Berry for valuable comments relating to the fractal Talbot effect.
 The author expresses his sincere thanks to Miss Pipa, the site administrator of  Russian Quantum Portal, for developing and writing a program calculating the density distribution of the wave function at scattering particles on $N$-slit grating. The program draws also Bohmian trajectories outgoing from the slits.
\end{acknowledgments}





\end{document}